\documentclass[aps,10pt,american,prl,twocolumn,nofootinbib,groupedaddress, floatfix, superscriptaddress,amsmath,amssymb,longbibliography,citeautoscript]{revtex4-2}

\usepackage{graphicx}
\usepackage{soul}
\graphicspath{{figs/}} 

\usepackage{dcolumn}
\usepackage{bm}
\usepackage{color}
\usepackage{enumitem}
\usepackage{float}

\definecolor{BLACK}{gray}{0}
\definecolor{WHITE}{gray}{1}
\definecolor{RED}{rgb}{1,0,0}
\definecolor{GREEN}{rgb}{0,1,0}
\definecolor{BLUE}{rgb}{0,0,1}
\definecolor{CYAN}{cmyk}{1,0,0,0}
\definecolor{MAGENTA}{cmyk}{0,1,0,0}
\definecolor{YELLOW}{cmyk}{0,0,1,0}
\definecolor{armygreen}{rgb}{0.55, 0.73, 0.0}

\newcommand{\ourNu}{2.609(7)} 
\newcommand{\ourNuNoError}{2.609}

\begin{document}

\title{Scaling collapse of longitudinal conductance near the integer quantum Hall transition}

\author{Elizabeth J. Dresselhaus}
\affiliation{Department of Physics, University of California, Berkeley, California 94720, USA}

\author{Bj\"orn Sbierski}
\affiliation{Department of Physics and Arnold Sommerfeld Center for Theoretical
Physics, Ludwig-Maximilians-Universit\"at M\"unchen, Theresienstr.~37,
80333 Munich, Germany }
\affiliation{Munich Center for Quantum Science and Technology (MCQST), 80799 Munich, Germany
}

\author{Ilya A. Gruzberg}
\affiliation{Ohio State University, Department of Physics, 191 West Woodruff Ave, Columbus OH, 43210, USA}

\date{\today}

\begin{abstract}

Within the mature field of Anderson transitions, the critical properties of the integer quantum Hall transition still pose a significant challenge. Numerical studies of the transition suffer from strong corrections to scaling for most observables. In this work, we suggest to overcome this problem by using the longitudinal conductance $g$ of the network model as the scaling observable, which we compute for system sizes nearly two orders of magnitude larger than in previous studies. We show numerically that the sizeable corrections to scaling of $g$ can be accounted for in a remarkably simple form which leads to an excellent scaling collapse. Surprisingly, the scaling function turns out to be indistinguishable from a Gaussian. We propose a cost-function-based approach, and estimate $\nu=\ourNu$ for the localization length exponent, consistent with previous results, but considerably more precise than in most works on this problem. Extending previous approaches for Hamiltonian models, we also confirm our finding using integrated conductance as a scaling variable.
\end{abstract}

\maketitle


{\it Introduction.}
In the field of critical phenomena, the renormalization group framework explains the significance of universal critical exponents and provides recipes to calculate them from experimental or numerical data \cite{Cardy-Scaling-1996}. One such method uses finite-size scaling and allows us to determine the exponent $\nu$, which governs the divergence of the emergent length scale (usually the correlation length)
\begin{equation}
\xi \sim |x|^{-\nu}.      
\end{equation}
Here, the control parameter $x$ is assumed to be in the vicinity of its critical value $x_c=0$. Finite-size scaling considers a (preferably) dimensionless observable $F$ for various $x$ in a system that has a finite extent $L$ in one or more directions. If $L$ exceeds all microscopic length scales, and sufficiently close to the fixed point, $F$ can only depend on the dimensionless ratio $L/\xi$, or  
\begin{equation}
F(x,L) = F(L^{1/\nu}x). \label{eq:F_scaling}      
\end{equation}

When this single-parameter scaling ansatz is valid, we can plot $F(x,L)$ against $z = L^{1/\nu}x$, 
and determine the value of the critical exponent $\nu$ as the number that gives the best collapse of the data. Beyond its simplicity, the power of this approach rests in the fact that the scaling function $F(z)$ does not need to be known a priori or expanded around $z=0$, but is obtained as a byproduct. 

Here we are concerned with critical properties of numerical models for the non-interacting integer quantum Hall transition (IQHT)~\cite{Huckestein1995,Kramer-Random-2005,EversMirlin:review}.
The divergent length scale $\xi$ is the localization length of single-particle wavefunctions. Chalker and Coddington (CC) proposed a simple network model\cite{Chalker-Percolation-1988} for the transition and analyzed the dimensionless quasi-1D Lyapunov exponents for systems of varying width $L$. Using the scaling ansatz~\eqref{eq:F_scaling} they obtained $\nu\simeq 2.5 \pm 0.5$. Subsequent studies also relied on the ansatz~\eqref{eq:F_scaling} and obtained $\nu$ in the range $2.3$--$2.4$ with error bars $\gtrsim 0.03$, see the review~[\onlinecite{Slevin-Finite-2012}] and references therein.

This state of affairs changed drastically when Slevin and Ohtsuki~\cite{Slevin-Critical-2009} reconsidered scaling in the CC model with refined numerical accuracy and found that their data could not be fit to the ansatz~\eqref{eq:F_scaling} due to strong corrections to scaling coming from irrelevant variables. In this case one has to add the least-irrelevant scaling variable with exponent $y<0$ as an argument in the scaling function. To leading order in $x$, this results in the ansatz 
\begin{equation}
F(x,L) = F(L^{1/\nu}x,L^{-y}). \label{eq:g_scaling_irrelevant}      
\end{equation}

This modification makes the use of a simple scaling collapse with $L^{1/\nu}x$ impossible. 
Instead, one must expand the right-hand-side of Eq.~\eqref{eq:g_scaling_irrelevant} as a polynomial, and determine a large number of unknown parameters from tedious least-squares fitting~\cite{Obuse-Finite-2012,Amado-Numerical-2011,Nuding-Localization-2015}. It is evident that a wide range of system sizes $L$ necessitates the use of polynomials of sufficiently high order, increasing the number of fitting parameters and partially counteracting the desired gain in accuracy. Thus, while Ref.~[\onlinecite{Slevin-Critical-2009}] reported the value $\nu = 2.593\, [2.587,2.598]$, most subsequent papers reported considerably larger error bars, as well as some scatter in the values for $\nu$  ($\nu=2.56-2.62$)~\cite{Obuse-Conformal-2010,Amado-Numerical-2011,Fulga2011a,Obuse-Finite-2012,Nuding-Localization-2015}. Also, no consensus has been reached on the value of $|y|$ besides that it likely is much smaller than unity. 

In spite of intensive efforts, more than a decade after the recognition of the importance of irrelevant corrections, the IQHT still evades full understanding. In contrast, for other Anderson transitions $\nu$ is typically known to three digit accuracy~\cite{EversMirlin:review}. 

In fact, the situation is even more severe: an increasing number of studies has questioned the very nature of the IQHT as a conventional localization transition with well-defined universal critical exponents. Notable examples include studies of different continuum models and lattice models~\cite{Zhu-Localization-length-2019, Puschmann-Integer-2019}, Dirac fermions~\cite{Sbierski2020c}, network models with two-channels~\cite{Lee-Chalker-PRL1994, Dresselhaus-Sbierski-Gruzberg-2021}, random geometry~\cite{Gruzberg2017,Klumper2019}, and models with dissipation~\cite{Beck2020}. 
Also, various versions of the Wess-Zumino model were proposed as analytical theories of the IQHT~\cite{Bhaseen-Towards-2000, Tsvelik-incollection, Tsvelik-Evidence-2007}, culminating in the recent proposal of a conformal field theory with only marginal perturbations ($\nu=\infty$)~\cite{Bondesan-Gaussian-2017,Zirnbauer2019, Zirnbauer-Marginal-CFT-perturbations-2021,Dresselhaus-Sbierski-Gruzberg-2021}. As a consequence of insufficient accuracy in numerical results, there is currently no consensus on any of the above conjectured deviations from the standard scaling scenario. 

There are numerical studies where scaling variables with no irrelevant corrections to scaling were observed: the scattering-matrix based variable~\cite{Fulga2011a}, the number of conducting states~\cite{Zhu-Localization-length-2019}, and the curvature of Lyapunov exponents at $x=0$~\cite{Slevin-Finite-2012}.
However, since numerically accessible system sizes are limited in these approaches, the resulting values of $\nu$ have to be considered cautiously.

\begin{figure}
\centering
\includegraphics{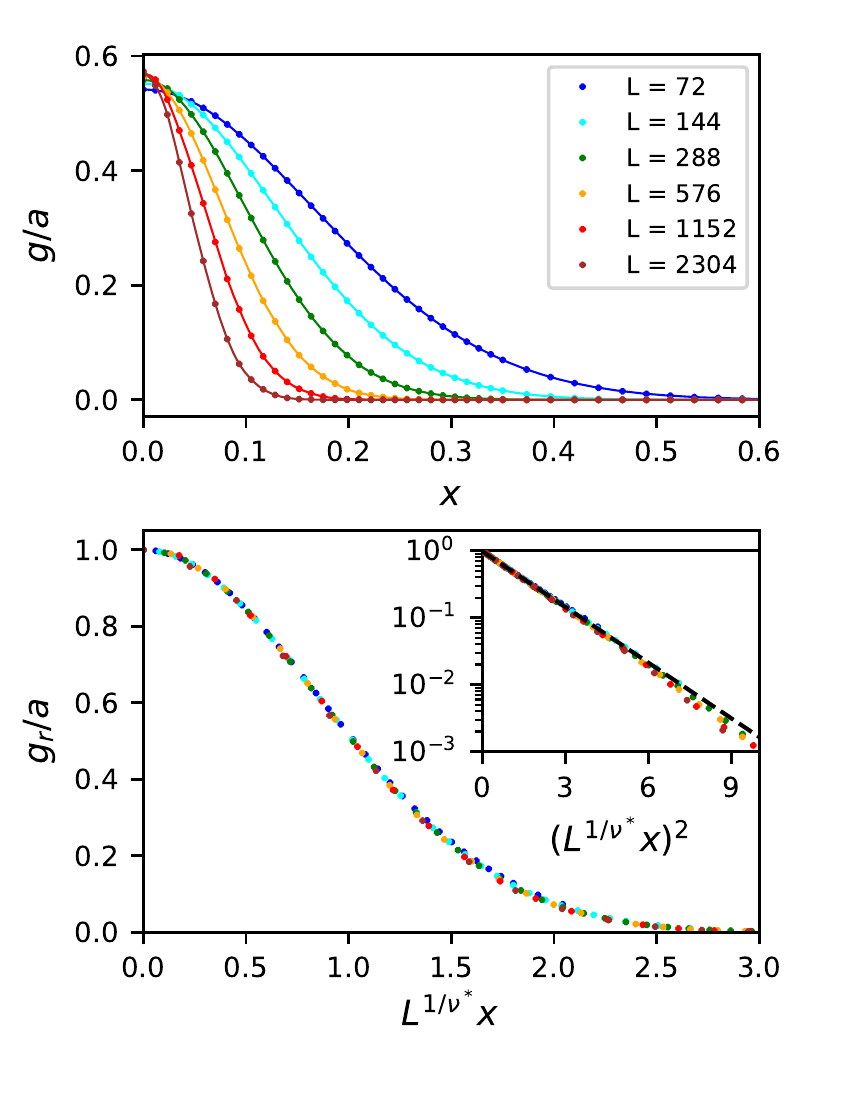}
\caption{Top: the mean longitudinal conductance~$g(x,L)$ for CC networks of $ L \times aL$ nodes with aspect ratio $a = 1$. The solid lines represent cubic-spline interpolations of the data (dots). Bottom: Scaling collapse of the the rescaled mean conductance, Eq.~\eqref{eq:g_r_def}, for $\nu^* = \ourNuNoError$ with the restriction $x \leq x_\text{max}=0.4$. The inset shows the same data vs.~$(L^{1/\nu^\star}x)^2$ on a logarithmic scale. The resulting straight line (dashed line is a guide to the eye) indicates that the scaling function is a Gaussian for not-too-large arguments. \label{fig:scaling_collapse_CC1}}
\end{figure}

In this work, we consider the longitudinal Landauer conductance of standard network models as the finite-size scaling variable, $F=g(x,L)$. Unlike the quasi-1D Lyapunov exponent, the longitudinal conductance, which is defined for two-dimensional systems, can be and has been measured in experiment~\citep{Cobden-Kogan-conductance_distribution_experiment}. However, in numerical simulations it also suffers from irrelevant contributions, see Fig.~\ref{fig:scaling_collapse_CC1} (top).
 
As our main idea, we demonstrate an empirical ansatz for a rescaled conductance $g_r(x,L)$ which to very high accuracy fulfills the standard single-parameter scaling without any observable irrelevant corrections. This insight allows us to faithfully re-introduce a scaling collapse analysis for the IQHT, involving data for system sizes varying by a factor of 32, see Fig.~\ref{fig:scaling_collapse_CC1} (bottom). The quality of the collapse is reminiscent of long-established scaling behavior found in classical statistical mechanics! In addition, we find the scaling function to be a simple Gaussian for not too large arguments.

We propose a cost-function approach to quantify and automate the search for the critical exponent $\nu$ from the best scaling collapse. For technical reasons, we change to the rescaled median conductance for this analysis. Taking into account variations in (i) minimal system size $L_\text{min}$, (ii) maximal tuning parameter $x_\text{max}$, and (iii) lower conductance cutoff, we obtain  $\nu=\ourNu$. We also report consistent results from the finite size scaling of the $x$-integrated median conductance. 

{\it Rescaled conductance and scaling collapse.}
We consider the single-channel CC network model~\cite{Chalker-Percolation-1988} in a rectangular geometry with $L \times aL$ nodes, and periodic boundary conditions in the direction with length $aL$. For each disorder realization, the Landauer conductance in the direction with length $L$ is obtained from the scattering matrix computed efficiently with the numerical method described in our previous work~\citep{Dresselhaus-Sbierski-Gruzberg-2021}. The conductance distribution depends on a parameter $x$ that drives the IQHT, with $x$ and $-x$ equivalent by symmetry of the model, and $x_c=0$. For square samples $(a=1)$, distributions of the critical ($x=0$) Landauer conductance take on a characteristic non-Gaussian shape~\cite{Kramer-Random-2005,EversMirlin:review}, while at larger $x$ or $L$ the distributions broaden even more and develop long tails, see appendix. We collected $N \simeq 10000$ disorder realizations for each system size.

The disorder-averaged (mean) conductance~$g(x,L)$ is shown by dots in Fig.~\ref{fig:scaling_collapse_CC1} (top). The data is qualitatively similar to that of dimensionless Lyapunov exponents in the apparent lack of a unique crossing point of different $L$-traces. In the standard scaling picture this implies the importance of an irrelevant scaling field $u_1(x)$ in the ansatz $g(x,L)=g(L^{1/\nu}u_0(x),L^{-y}u_1(x))$, where $u_0(x)$ is the relevant scaling field. While $u_{0,1}(x)$ are generally unknown, their leading order behavior for $x \ll 1$  is $u_0(x)\sim x$ and $u_1(x) = \mathrm{const}$, resulting in $g(x,L) = g(L^{1/\nu}x,L^{-y})$, c.f.~Eq.~\eqref{eq:g_scaling_irrelevant}. 

In what follows, we will demonstrate that the conductance exhibits the factorized form
\begin{equation}
g(L^{1/\nu}x,L^{-y})=g_0(L^{1/\nu}x)g_1(L^{-y}). 
\label{eq:conjecture}
\end{equation}
If the ansatz~\eqref{eq:conjecture} holds, the rescaled conductance, which we define as 
\begin{equation}
g_r(x,L)=g(x,L)/g(x=0,L) \label{eq:g_r_def}
\end{equation}
will have no irrelevant contribution and should show a scaling collapse,
\begin{equation}
g_r(x,L)=g_0(L^{1/\nu}x)/g_0(0)=g_r(xL^{1/\nu}). \label{eq:g_r_scaling}
\end{equation}

In Fig.~\ref{fig:scaling_collapse_CC1} (bottom) we plot $g_r(xL^{1/\nu})$ for a range of sizes $L$ varying by a factor of 32. The perfect collapse for $\nu=\nu^\star=\ourNuNoError$ shows one of our main results: the rescaled conductance is a scaling observable without irrelevant contributions. 

To investigate the functional form of the scaling function which controls the conductance $g(x,L)$ in the large-$L$ limit when irrelevant corrections are practically absent, 
we plot $g_r$ as a function of $(xL^{1/\nu^\star})^2$ on a logarithmic scale, see the inset in Fig.~\ref{fig:scaling_collapse_CC1} (bottom). The resulting straight line indicates that the scaling function is a simple Gaussian. This surprising result places strong constraints on putative analytical  theories of the IQHT~\cite{Zirnbauer-Marginal-CFT-perturbations-2021}. Notice that we do not expect the scaling function to remain Gaussian for large values of the argument, since this corresponds to the localized phase with $g \sim e^{-L/\xi}$. 

In order to quantitatively analyze the quality of the scaling collapse beyond visual inspection and to take into account errors of the raw data, we proceed with a cost-function analysis.
Due to the aforementioned non-Gaussian shape for conductance distributions for large $x$ and $L$, the mean and its error are difficult to estimate with a limited number of realizations. Therefore, we prefer to base the subsequent analysis on the \textit{median} conductance, denoted by $\gamma(x, L)$. Our analysis supports the expectation that the critical exponent~$\nu$ is universal and thus can be found from either $g$ or $\gamma$. This is not true for the scaling function, which shows slight deviations from a Gaussian for $\gamma$. To find the error of the median we use the asymptotic variance formula, $\sigma^2 = [4N P^2(\gamma)]^{-1}$,
where $P(\gamma)$ is the spline-interpolated probability distribution of $g$ evaluated at $g = \gamma$ and $N$ is the number of disorder realizations. 

We arrange the system sizes in increasing order $L_1 < \ldots < L_{N_L}$. For each system size $L_i$, we define $f_i(z)=\gamma_r(L_i,x=z L_i^{-1/\nu})$ using a cubic spline interpolation of the rescaled median $\gamma_r(L_i, x)$ with respect to $x$ for all $x$ that fulfill (i) $x \leq x_\text{max}$ and (ii) $\gamma_r(L_i, x) \geq \gamma_{r,\text{min}}$. Restriction (i) is required by the approximation $u_0 \sim x$ in Eq.~\eqref{eq:g_scaling_irrelevant}, while (ii) allows us to exclude too heavily skewed conductance distributions for which even the estimate of the median might become problematic. Both conditions together define a range $z \leq z^\text{max}_i$. We compute the error $\delta_i(z)$ of $f_i(z)$ by error propagation using the error $\sigma(L_i,x)$ of $\gamma(L_i,x)$.

\begin{figure}[t]
\centering
\includegraphics{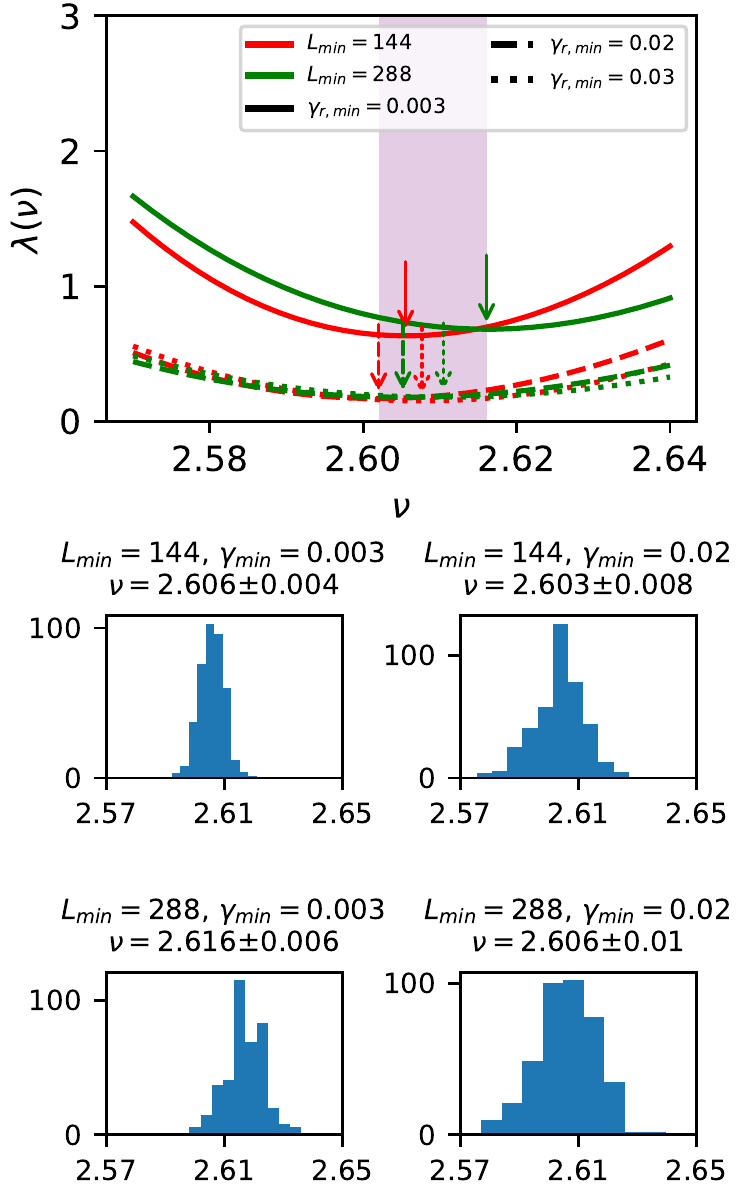}
\caption{Top: The cost function $\lambda(\nu)$. Minimum $\nu^*$ estimates the critical exponent of the IQHT. We restrict to data points where $\gamma_r \geq \gamma_\text{r, min}, L \geq L_\text{min}$ for several choices of $\gamma_\text{r, min}, L_\text{min}$. In this analysis we fix $x_\text{max} = 0.7$. The shaded region spans the minima for all combinations of these parameters that result in minimum cost $ \lambda_\text{min} < 1$. Note that we exclude $L = 72$ in this analysis but include it in the scaling collapse of Fig.~\ref{fig:scaling_collapse_CC1}. Bottom: Histograms for the best $\nu$ obtained from the synthetic data method described in the main text below Eq.~\eqref{eq:nu_CC1}.
\label{fig:cost_function}}
\end{figure}

Single-parameter scaling means that, for appropriately chosen $\nu$, all $f_i(z)$ should collapse onto a single curve defining the scaling function. We thus assert that the random variable $f_{i}-f_{j}$ is Gaussian distributed with zero mean and variance $\delta_i^2+\delta_j^2$. For $i<j$ we can evaluate the collapse quality for $z\in [0,z^\text{max}_{i,j}]$ with $z^\text{max}_{i,j} = \text{min}(z^\text{max}_i, z^\text{max}_j)$, 
using the following definition of a cost function,
\begin{equation}
    \lambda(\nu) =  \frac{1}{N_P} \sum_{i<j} \frac{1}{z^\text{max}_{i,j}}\int_{0}^{z^\text{max}_{i,j}} \frac{|f_i(z) - f_j(z)|^2}{\delta^2_i(z)+\delta_j^2(z)} dz,
    \label{eq:cost_function}
\end{equation}
where the summation is over the $N_P = N_L(N_L-1)/2$ pairs of distinct system sizes. The best $\nu$ is the one that minimizes the cost function, and the scaling hypothesis can only be accepted if the minimum of the cost function is smaller than unity. 

In Fig.~\ref{fig:cost_function} (top), we show $\lambda(\nu)$ for $x_\text{max} = 0.7$, $\gamma_{r,\text{min}}=0.003$, and for system sizes $L \geq L_\text{min} = 144$ (red solid line). The optimal $\nu$ falls close to $\ourNuNoError$ where $\lambda \leq 1$ indicates an excellent collapse. Choosing $\gamma_{r,\text{min}} = 0.02$, and $0.03$, we obtain the red dashed and dotted curves, respectively. Their minima are indicated by arrows. Likewise, consistent results are obtained for doubling $L_\text{min}$, see the green lines. We exclude minima if $\lambda(\nu_\text{min}(\gamma_{r,\text{min}}, L_\text{min})) > 1$. Taking the spread of all included minima (the shaded region in Fig.~\ref{fig:cost_function}), we obtain
\begin{equation}
\nu = \ourNu, 
\label{eq:nu_CC1}
\end{equation}
which is in the previously reported range but with the error that is smaller than in most previous studies. 

\begin{figure}
\centering
\includegraphics{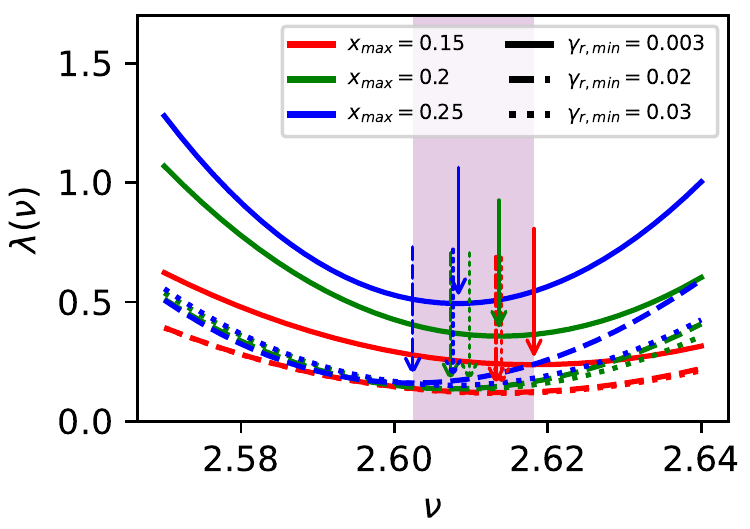}
\caption{The cost-function analysis for $L_\text{min} = 144$, restricted to data points where $\gamma_r \geq \gamma_\text{r, min}$ and $x \leq x_\text{max}$. The values of $\gamma_\text{r,min}$ in Fig. \ref{fig:cost_function} are used in tandem with $x_\text{max} = 0.15, 0.2, 0.25$. The shaded region spans the minima for all combinations of these parameters that result in minimum cost $ \lambda_\text{min} < 1$. Results are consistent with Eq.~\eqref{eq:nu_CC1}. 
\label{fig:cost_function_check}}
\end{figure}

Consistent results for $\nu$ are obtained from the synthetic data method. Using the input data for $\gamma_r$ and its error, we generate 400 sets of synthetic data on which to repeat the cost-function analysis of Eq.~\eqref{eq:cost_function}. The bottom panel of Fig.~\ref{fig:cost_function} shows the histograms of the best values of $\nu$ obtained this way, for $L_\text{min}$ as above and $\gamma_\text{min}= 0.003, 0.02$. The means for all histograms fall within the range indicated in Eq.~\eqref{eq:nu_CC1}, and their standard deviation is on the order of $0.01$.

We repeat the cost-function analysis for $L_\text{min} = 144$ with three more choices of $x_\text{max} = 0.15, 0.2, 0.25$, see Fig.~\ref{fig:cost_function_check}. The resulting spread of the minima $\lambda_\text{min}$ (the shaded region) is nearly identical to that of Fig.~\ref{fig:cost_function}, and includes the estimate of Eq.~\eqref{eq:nu_CC1}.

{\it Integrated Conductance.}
Once a scaling observable is thought to obey the single-parameter scaling like $g_r(x,L)=g_r(xL^{1/\nu})$ above, an alternative option for the evaluation of $\nu$ is via the $x$-integrated observable 
\begin{equation}
    \tilde{g}_r(L) = \int_{0}^{\infty} g_r(x,L)dx \sim L^{-1/\nu},
    \label{eq:tilde_gr}
\end{equation}
where the predicted scaling behavior on the right hand side follows from a substitution. Sidestepping the scaling collapse check, this approach however crucially requires that the system sizes are large enough so that the integrand is negligible in regions where higher-order-in-$x$ corrections to the scaling function become important (c.f.~the discussion of $x_\text{max}$ above). We thus chose $L \geq 144$. 

\begin{figure}
\centering
\includegraphics{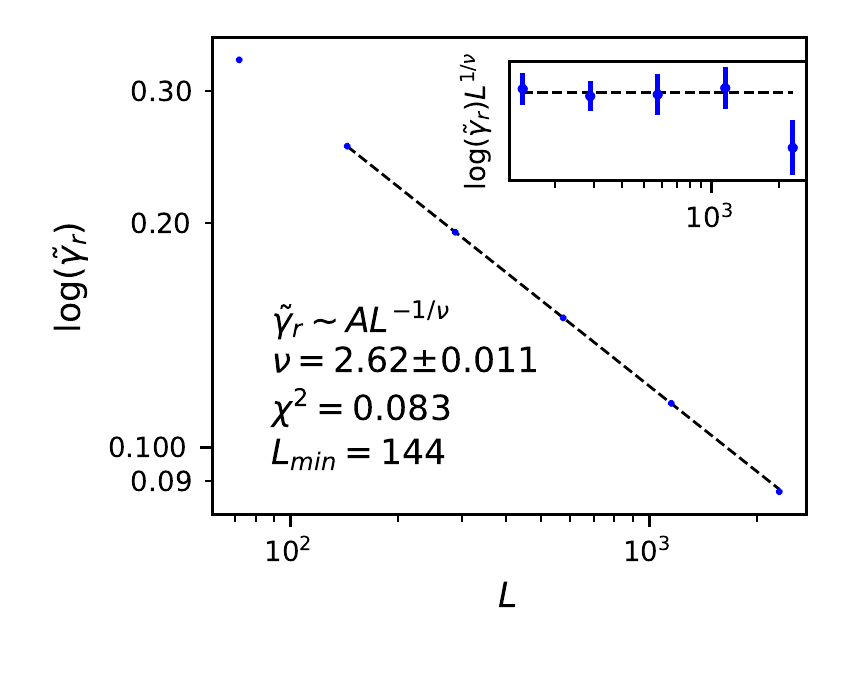}
\caption{Scaling of the $x$-integrated rescaled median conductance of the CC model. A power-law fit for $L\geq 144$ is denoted by a dashed line, the fit parameters are given in the legend. The inset shows the data in a tilted version for better visualization.
\label{fig:cc1_intg}}
\end{figure}

To compensate for under-sampling the rare regions in the tails of the conductance distributions discussed earlier, we integrate the median rescaled conductance $\gamma_r$. We generate 200 artificial data sets that lie within the error bars of $\gamma_r(x, L)$, apply cubic spline interpolation and integration, and estimate the mean and standard deviation of the 200 values to be $\tilde{\gamma}_r(L)$ and its associated error. The results are shown in Fig.~\ref{fig:cc1_intg}. The fit to  Eq.~\eqref{eq:tilde_gr} yields $\nu$ consistent with the result~\eqref{eq:nu_CC1}. 

We remark that a similar integrated scaling observable for the IQHT was proposed and studied previously in Refs.~[\onlinecite{Zhu-Localization-length-2019}] and~[\onlinecite{Bhatt-original}]. These works considered the number of conducting eigenstates of lattice- and continuum Hamiltonian models, but did not include a rescaling of the energy-resolved data which thus may have had irrelevant contributions that changed the right-hand-side of Eq.~\eqref{eq:tilde_gr} and affected the extracted value of $\nu$. Moreover, the use of the exact diagonalization limited the available system sizes, which might also explain a significantly smaller result~\cite{Zhu-Localization-length-2019} for $\nu=2.48(2)$. 

{\it Conclusions and outlook.}
To summarize, in the context of the IQHT, we have identified the rescaled longitudinal Landauer conductance of the CC network model as a promising scaling observable. It satisfies the single-parameter scaling and exhibits no detectable irrelevant corrections. This allows us to sidestep the tedious polynomial expansion of scaling functions that is commonly done in the literature, and apply the intuitive and straightforward scaling collapse for a wide range of system sizes with $L_\text{max}/L_\text{min} = 32$. We devise a cost-function approach to quantify the quality of the collapse and automate the extraction of the critical exponent. Our result $\nu=\ourNu$ is consistent with previous estimates in the literature but with higher accuracy than most studies.

This strategy also gives access to the scaling function of the mean conductance, which takes a surprisingly simple Gaussian form; this will inform further analytical studies on the subject. Inspired by previous works~\cite{Bhatt-original,Zhu-Localization-length-2019} on the total number of conducting states in Hamiltonian systems, we also consider the integral of the rescaled conductance which sidesteps the necessity for a scaling collapse and allows for even simpler fitting.

Our findings might also offer a new approach to experimental studies of the IQHT, even though critical properties in real samples may be strongly affected by the presence of electron-electron interactions. The latest scaling analysis of experimental data is based on the slope of the Hall conductivity with respect to the magnetic field~\cite{LiPRL2005, Li_quantum_hall_scaling_experiment2009, LiPRB2010_crossoverExperiment}. However, the longitudinal conductance peak as a function of the magnetic field is routinely measured as well. While earlier studies have already used the peak width as a scaling observable~\cite{Huckestein1995}, we propose to consider the full shape of the peak measured for various sample sizes below the phase coherence length, a regime reached in Ref.~[\onlinecite{LiPRB2010_crossoverExperiment}]. After centering to the respective peak maximum and performing rescaling as in Eq.~\eqref{eq:g_r_def}, a scaling collapse could be achieved. In order to get a good estimate of the mean or median conductance, it will be essential to revisit the issue of the full conductance distributions, see Ref.~[\onlinecite{Cobden-Kogan-conductance_distribution_experiment}] for previous experimental work.

We emphasize that the results presented above do not settle the question about the validity of the marginal scaling scenario~\cite{Bondesan-Gaussian-2017,Zirnbauer2019, Zirnbauer-Marginal-CFT-perturbations-2021} according to which the exponent $\nu$ would be scale- and model-dependent~\cite{Dresselhaus-Sbierski-Gruzberg-2021}. While our results lower the upper bound for a possible scale dependence of $\nu$, it is conceivable that it might be revealed in future studies reaching even larger scales or accuracies. We also suggest to attempt a scaling collapse of the rescaled conductance distribution as a whole. Further, it would be interesting to test if a rescaling procedure can also eliminate the irrelevant contributions to the quasi-1D Lyapunov exponents, which is a standard scaling observable in the literature~\cite{Slevin-Critical-2009}. 

Regarding a possible model dependence of $\nu$, recent studies indeed seem to point in this direction~\cite{Zhu-Localization-length-2019, Puschmann-Integer-2019, Sbierski2020c, Lee-Chalker-PRL1994, Dresselhaus-Sbierski-Gruzberg-2021, Klumper2019,Beck2020}. As a demonstration of the scaling-collapse method applied for an alternative model, we consider the two-channel CC network model (CC2)~\cite{Lee-Chalker-PRB1994,Lee-Chalker-PRL1994} in the appendix. We show that the rescaled conductance collapses for $\nu_\mathrm{CC2} = 3.8(1)$, significantly different from the exponent for the (single-channel) CC case but in line with recent results using a different method~\cite{Dresselhaus-Sbierski-Gruzberg-2021}.

\vspace{5mm}
\begin{acknowledgments}
We acknowledge useful discussions with Ravindra Bhatt, Nils Niggemann, Sasha Mirlin, Ferdinand Evers, Tomi Ohtsuki and Keith Slevin. Computations were performed at the Lawrencium cluster at Lawrence Berkeley National Lab. EJD received financial support from the Graduate Research Fellowship program, USA, NSF DGE 1752814. BS acknowledges financial support by the German National Academy of Sciences Leopoldina through Grant Number LPDR 2021-01, by a MCQST-START fellowship and by the Munich Quantum Valley, which is supported by the Bavarian state government with funds from the Hightech Agenda Bayern Plus. 

\end{acknowledgments}

\appendix

\section{Appendix}

\begin{figure}[h]
\centering
\includegraphics[scale = .9]{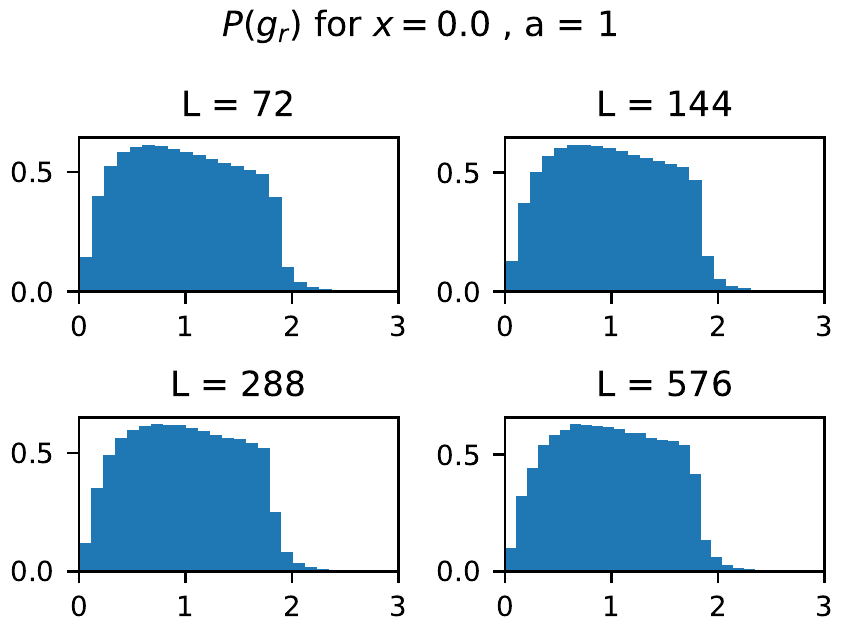}
\includegraphics[scale = .9]{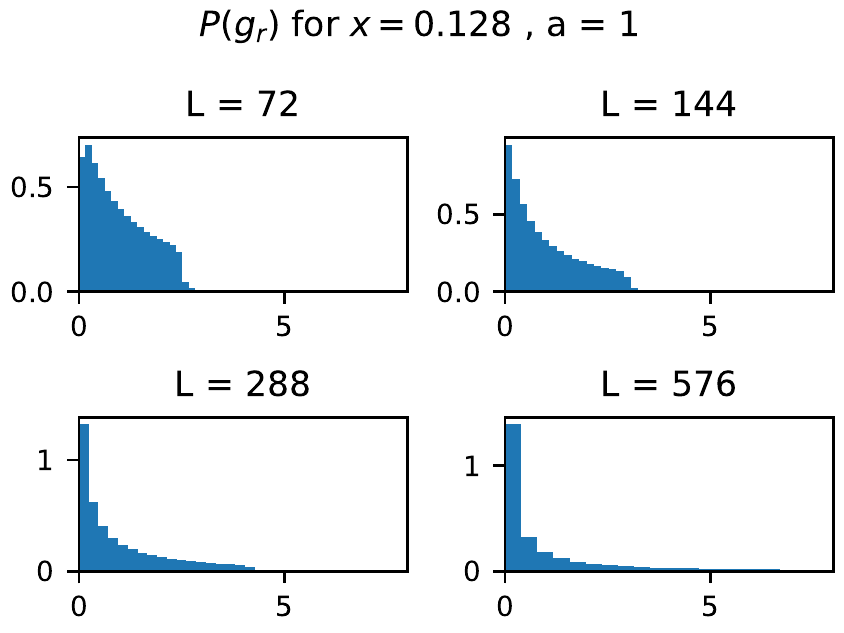}
\caption{Top: Critical conductance distribution for square CC network model of size $L$.
Bottom: Conductance distribution at $x = 0.128$ (non- critical) for square CC network model of size $L$.}
\label{fig:a_1}
\end{figure}

{\it Conductance distributions.} 
We present conductance distributions for $N \simeq 10000$ square CC network models at fixed $x=0$ and $x=0.128$, see Fig.~\ref{fig:a_1}. Away from criticality ($x>0$), the distribution becomes increasingly skewed with larger system size.The shape of these distributions motivates the use of the median rather than the mean for the scaling collapse analysis of the main text.

\clearpage 
{\it Two-channel CC network model.}
The two-channel generalization of the CC model (CC2) was first studied in Refs.~\cite{Lee-Chalker-PRB1994,Lee-Chalker-PRL1994}. The present authors previously studied this model and its phase diagram in Ref.~\citep{Dresselhaus-Sbierski-Gruzberg-2021} and we refer to this reference for a detailed discussion. Here we limit ourselves to the diagonal $(x_c+x,x_c+x)$ in the phase diagram of the CC2 which is spanned by the tuning parameters $(x_a,x_b)$ of the individual layers and $x_c=0.227$. At the point $(x_c, x_c)$, the critical line of the phase diagram intersects the diagonal~\citep{Dresselhaus-Sbierski-Gruzberg-2021}. As for the CC, we obtain the rescaled longitudinal conductance for the CC2 at sizes $L = 288, 576, 1152, 2304$ ($N \simeq 1000$ realizations) at various $x>0$ and examine the scaling collapse.

\begin{figure}[h]
    \centering
    \includegraphics{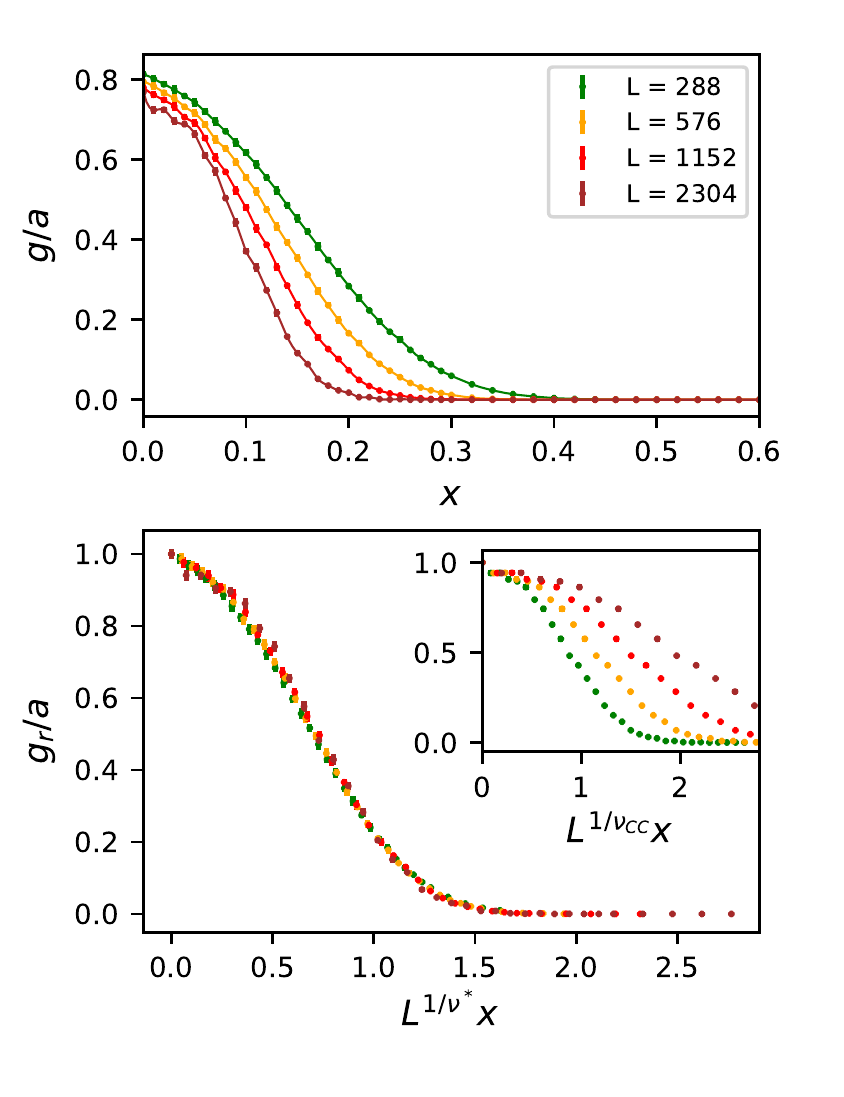}
    \caption{Top: Mean longitudinal conductance $g(x_c + x,x_c + x, L$) of square CC2 networks. Solid lines represent cubic-spline interpolations of the data (dots). Bottom: The inset shows the rescaled longitudinal conductance of various systems of size $L$ and its failure to collapse with the critical exponent of the single-channel CC network, $\nu_{CC} = 2.6$. This suggests the CC2 has a different critical exponent. The main panel shows a reasonable collapse for $\nu^* = 3.9$ with the restriction $x < 0.4$.}
    \label{fig:cc2_collapse}
\end{figure}

\begin{figure}[h]
    \centering
    \includegraphics{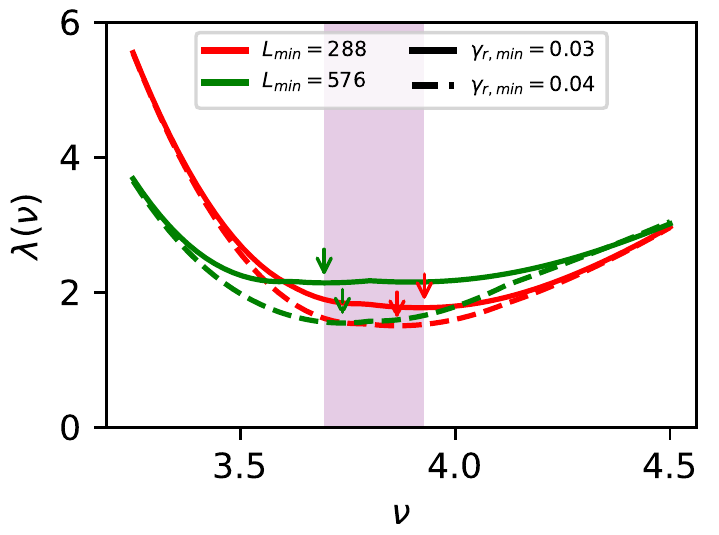}
    \caption{The cost function $\lambda(\nu)$ estimates the critical exponent for the IQHT in the CC2 model. We restrict to data points where $\gamma_r \geq \gamma_{r, min}, L \geq L_{min}$ for several choices of $\gamma_{r, min}, L_{min}$. The shaded region spans the minima for all combinations of these parameters shown. Note that while no choice gives $\lambda(\nu)_{min} \leq 1$ in this case, however, the cost does show a clear minimum with $\lambda(\nu)_{min} \simeq 1$.}
    \label{fig:cc2_cost}
\end{figure}

In Figs.~\ref{fig:cc2_collapse} and~\ref{fig:cc2_cost}, we repeat the analysis of the main text. We find $\nu_{CC2} = 3.8(1)$, substantially different from our estimates for the single-channel case. Our finding also agrees with our result for the same critical point using an alternative scaling variable~\citep{Dresselhaus-Sbierski-Gruzberg-2021}, $\nu = 3.90(5)$.

\newpage
\clearpage
\bibliography{library}

\end{document}